\documentstyle[12pt,epsfig]{article}

\begingroup\makeatletter\ifx\SetFigFont\undefined
\def\x#1#2#3#4#5#6#7\relax{\def\x{#1#2#3#4#5#6}}%
\expandafter\x\fmtname xxxxxx\relax \def\y{splain}%
\ifx\x\y   
\gdef\SetFigFont#1#2#3{%
  \ifnum #1<17\tiny\else \ifnum #1<20\small\else
  \ifnum #1<24\normalsize\else \ifnum #1<29\large\else
  \ifnum #1<34\Large\else \ifnum #1<41\LARGE\else
     \huge\fi\fi\fi\fi\fi\fi
  \csname #3\endcsname}%
\else
\gdef\SetFigFont#1#2#3{\begingroup
  \count@#1\relax \ifnum 25<\count@\count@25\fi
  \def\x{\endgroup\@setsize\SetFigFont{#2pt}}%
  \expandafter\x
    \csname \romannumeral\the\count@ pt\expandafter\endcsname
    \csname @\romannumeral\the\count@ pt\endcsname
  \csname #3\endcsname}%
\fi
\fi\endgroup

\advance \topmargin by -\headheight
\advance \topmargin by -\headsep

\evensidemargin \oddsidemargin
\begin{document}

\newpage
\pagestyle{empty}

\begin{flushright}
gr-qc/0104005\\
$~$ \\
April 2001
\end{flushright}

\begin{centering}

\baselineskip 21pt plus 0.2pt minus 0.2pt

\bigskip
{\Large {\bf Phenomenological description of space-time foam} }
 
\bigskip

\baselineskip 12pt plus 0.2pt minus 0.2pt

\bigskip
\bigskip
\bigskip
\bigskip

{\bf Giovanni AMELINO-CAMELIA}\\
\bigskip
Dipartimento di Fisica, Universit\'{a} ``La Sapienza", P.le Moro 2,
I-00185 Roma, Italy\\ 
\end{centering}
\vspace{1.5cm}
     
\centerline{\bf ABSTRACT }
\medskip

\begin{quote}

The expectation that it should not be possible to gain experimental insight
on the structure of space-time at Planckian distance scales 
has been recently challenged by several studies.
With respect to space-time fluctuations,
one of the conjectured features of quantum-gravity foam,
the experiments that have the best sensitivity are the ones
which were originally devised for searches of the classical-physics 
phenomenon of gravity waves. 
In experiments searching for classical gravity waves the presence
of space-time fluctuations would introduce a source of noise just
like the ordinary (non-gravitational) 
quantum properties of the photons composing the laser beam 
used in interferometry introduce a source of noise.
Earlier studies of the noise induced by quantum properties of
space-time have shown that certain simple pictures
of fluctuations of space-time occuring genuinely at the Planck scale
would lead to an observably large effect.
Experimentalists would benefit from the guidance of detailed
description of this noise, but quantum-gravity theories are not
yet developed to the point of allowing such detailed analysis
of physical processes. I propose a new phenomenological approach
to the description of foam-induced noise.

\end{quote}
\baselineskip 18pt

\vfill

\newpage
\pagenumbering{arabic}
\setcounter{page}{1}
\pagestyle{plain}
\baselineskip 12.5pt plus 0.2pt minus 0.2pt

\section{Introduction and summary}

Work done in the last two decades~\cite{cow,ehns}
and particularly over the last few 
years~\cite{venegwi,ahluexp,grbgac,gacgwi,kifu,gactp}
has corrected an old misconception that
it would not be possible to gain experimental insight 
on ``quantum gravity"
(the sought theory of the
interplay between general relativity and quantum mechanics)
and on the structure of space-time at Planckian distances.
The pessimistic expectations 
for ``quantum-gravity phenomenology"\footnote{In spite of the fact
that I am somewhat responsible~\cite{polonpap} for the fact that
the community refers to this field as ``quantum-gravity phenomenology",
I now believe that this is a rather inappropriate terminology.
Especially for a phenomenological (hopefully data-driven)
programme it is not proper to make any a priori dogmatic assumptions
about the outcome of the studies to be conducted, and the 
name quantum-gravity phenomenology could erroneously
suggest that we have experimental evidence indicating that a relatively
straightforward quantization of Einstein's gravity is realized in Nature
(while, as it is well known, there is no evidence of this type).
A better name for the field would have 
been ``Planck-length phenomenology"~\cite{grf00},
reflecting the fact that its objective is the investigation
of the structure of space-time at Planckian distance scales.
We also do not have any robust experimental evidence of new physics occuring
at those distance scales, but the question ``what is the structure of
space-time at Planckian distances?" is fully meaningful physically: it is
in a sense operatively well defined. (Moreover, the conceptual arguments
suggesting that new physical theories are required in the Planck-length
regime come from several complementary lines of analysis and appear to 
be robust.)}
that are described in traditional quantum-gravity reviews
basically rely on two simple observations.
First one observes that 
the interplay of general relativity and quantum mechanics
can be the dominant element in the analysis of a physical
context only if this context involves strong gravitational
forces and short distances.
This is a condition that was realized in the early stages
of evolution of the Universe, but 
we would not be able to realize similar
conditions experimentally.
The second observation is based on the fact that
in the contexts that we can study experimentally,
which only involve length scales much larger than the Planck 
length ($L_p \equiv \sqrt{\hbar G/c^3} \sim 10^{-33} cm$),
all effects induced by quantum gravity would be very small.
In fact, since $L_p$ is proportional to both the gravitational 
constant, $G$, and the Planck contant, $\hbar$, 
we expect that the magnitude
of these effects should be set by
some power of the ratio
between the Planck length and a characteristic length scale
of the process under investigation.

Even without any detailed analysis of the  
interplay between general relativity and 
quantum mechanics it should be clear that these
two observations cannot be sufficient for justifying the radical
assumption that there is no hope for quantum-gravity phenomenology.
In fact, for example,
similar arguments would apply also to
grandunified theories of the electroweak and strong interactions,
which predict large new effects for collision processes 
involving particles with wavelengths of the order 
of $10^{-30} cm$ (which, of course, are not available to us),
but predict only very small effects for the
type of processes we can access experimentally.
Yet experimentalists did manage to devise experiments
with very good sensitivity to grandunification predictions.
The probability of proton decay in grandunification
is extremely small,
since it is governed by
the fourth power of the ratio between the mass of the proton
and the grandunification scale,
but, in spite of such a strong suppression,
experimentalists are
managing to set significant bounds
on proton decay by keeping under observation
a large number of protons (so that the experiments
measure the probability that one among
many protons decayes).

As the analogy with proton decay in grandunification
suggests, quantum-gravity phenomenology is not impossible,
but it is of course extremely hard.
Opportunities to test experimentally
the nature of the interplay between
general relativity and quantum mechanics
remain extremely rare, but we have now a handful of 
proposals~\cite{cow,ehns,venegwi,ahluexp,grbgac,gacgwi,kifu,gactp}
which represent a significant step forward with respect to 
the expectations of not many years ago.

In this paper I focus on one of these
opportunities for quantum-gravity phenomenology: the study of
quantum space-time fluctuations using modern 
interferometers~\cite{gacgwi,bignapap,nggwi,emngwi,fordgwi}.
My main objective here is the proposal of a new phenomenological
approach to the description of quantum-gravity-induced interferometric
noise. This approach was recently sketched out in a short non-technical 
paper~\cite{gacgwi2}, and here I intend to elaborate on the proposal,
also discussing some of the technical points omitted in Ref.~\cite{gacgwi2}.

Since quantum-gravity phenomenology is still a relatively young
research subject, I start, in the next Section, with a brief
review of the status of this field.
I describe the main proposals, with emphasis on one of
these proposals which has recently enjoyed some success, in the sense
that it provided the first ever
instance of experimental data that appear to be a manifestation
of Planck-scale physics.
While this exciting development is not directly connected 
with the type of space-time-foam studies which are the main
focus of this Article, it appeared appropriate to discuss it in
some detail since it shows that the development of
quantum-gravity phenomenology has already encountered a 
non-empty set (one example)
of ``success stories", at least in the sense that the analysis of
certain present-day experimental paradoxes does appear to invite
considerations of the type that is natural in quantum-gravity
phenomenology. This in turn might provide encouragement to
colleagues involved in experiments relevant for my present
proposal of space-time-foam studies: having found already
one example in another quantum-gravity phenomenology
research line, one can look with more optimism at the possibility
of encountering a second example of ``quantum-gravity phenomenology
success story", and perhaps this second example will emerge
as we pursue the space-time-foam experimental programme here outlined.

In Section 3 I start developing my characterization of
space-time foam based on distance noise. 
I give an operative definition of space-time foam
and show that the sensitivity of modern interferometers could
be sufficient for detecting distance fluctuations occurring
genuinely at the Planck scale. I illustrate my point by showing
that the sensitivity achieved by modern interferometers
would, for example, be sufficient for detecting
the distance-noise level that corresponds to 
quantum fluctuations of the length of the arms of the interferometer
(of a few kilometers!) that are only of Planck-length magnitude and
occur with a frequency of one per Planck time.

In Section 4 I observe that (having established in Section 3 that
modern interferometers are potentially relevant for quantum-gravity
studies) it would now be necessary for the various theoretical 
approaches to the quantum-gravity problem to provide to the
experimentalists detailed models of the noise to be expected
in interferometers, but I also argue that the development
of these theoretical approaches is still too preliminary for
such ``physical" predictions. As a way to by-pass this limitation
of quantum-gravity theories, I then propose a new phenomenological
approach that describes directly 
quantum-gravity-induced distance noise.
My task is partly facilitated by the fact that
in order to guide interferometric studies
of foam it is only necessary to estimate
a relatively simple (single-variable) function:
the power spectrum of the distance noise~\cite{saulson,rwold}.
The quantum-gravity-induced strain noise should depend only
on its variable argument, the frequency $f$ at which observations
are made,
on the Planck length, 
on the speed-of-light constant $c$ ($c \simeq 3 {\cdot} 10^8 m/s$),
and, perhaps, on a length scale characterizing the properties
of the apparatus with respect to quantum gravity.
I find that within this conceptual framework
there is a compellingly-simple candidate
for a foam-induced ``white noise''
(noise with constant, $f$-independent, power spectrum).
More generally, I show that upon adopting a given
qualitative description of quantum distance fluctuations
({\it  e.g.}, ``white noise",  ``$f^{-1}$ 
noise", ``random-walk $f^{-2}$ noise"....)
one already has a lot of information on distance noise,
because dimensional analysis constrains very strongly 
the structure of the noise spectrum.
This leads me to consider a few natural candidates
for a phenomenological dscription of quantum-gravity-induced
distance noise, and, surprisingly, I find that over the
next few years experiments will put under scrutiny quite
a few of these candidates.

In Section~5 I discuss some urgent theory issues which
could benefit the development of the phenomenological
approach here proposed. The main point made in Section~4 is that
the information on distance noise needed by experimentalists
is such that with only a few qualitative guidelines from theory 
one can develop rather powerful phenomenological models.
However, the guidelines we can extract presently from 
quantum-gravity theories are very limited. In Section~5 I describe
a sort of ``wish list" describing the type of guidelines that
my new phenomenological approach most urgently needs from
theory. One important bit of qualitative information concerns
the structure of the limiting procedure, which of course must
be present in quantum gravity, by which a quantum space-time
approaches its corresponding classical space-time.
The outlook of model-building for distance noise would
be affected strongly by results indicating, for example,
that this classical-limit procedure must involve in some way
the ratio of the wavelengths of the particles versus
the Planck length (as one would expect in pictures in which the
same space-time appears to be ``quantum" to particles of
very short wavelength and appears to bee classical to particles
of very large wavelength, which basically average over the
short-distance structure of space-time).
Another important theory issue concerns the role of energy 
considerations in quantum gravity. 
While energy considerations are rather elementary
when the analysis is supported by a fixed background space-time,
the fact that quantum gravity cannot {\it a priori} 
rely on a background space-time 
renders energy considerations much more subtle~\cite{thooftnew}.
If however theoretical analyses eventually develop that capability
of providing some (even partial) information on the role of energy
considerations in quantum gravity, the phenomenological approach
I am proposing would acquire a very powerful tool
for discriminating between different distance-noise models.
 
After the description of a ``wish list for theory" given in Section~5,
in Section~6 I outline a sort of ``wish list for experiments",
emphasizing some key points that experimentalists planning
to contribute to this research programme should take into account.

Finally Section~7 is devoted to some closing remarks.

\section{Status of quantum-gravity phenomenology}

As mentioned, we finally have some, although still very
few, research lines attempting to gain experimental
information on the interplay between quantum mechanics and
general relativity.
Let us denominate ``quantum gravity" the, still unknown, theory
that describes this interplay.
There are clearly two qualitatively different features
that one can expect in quantum gravity:
(i) the presence of new particles ({\it e.g.} the graviton)
(ii) the presence of completely new phenomena~\cite{thooftnew},
not describable in terms of the propagation of a particle
in a background space-time,
related with the non-classicality of space-time.
For short I'll refer to (i) as ``particle-like effects"
or ``new particles propagating in a classical background
space-time" 
and to (ii) as ``non-particle-like effects"
or simply ``genuinely quantum space-time".

Of course the most exciting class of quantum-gravity phenomena
are the ones involving ``non-particle-like effects", the ones
related to the emergence of profoundly non-classical (possibly quantum)
properties of space-time, such as noncommutativity, dicreteness,
topology fluctuations.... 
General arguments~\cite{thooftnew,ahlu94grf,gacgrf98}
as well as specific proposals based on space-time
discreteness or 
noncommutativity~\cite{spinnet,kpoin,gacmaj,bilal,stringnonco}
provide motivation for the exploration of 
this ``non-particle" possibility.
I start this short review from
those experiments which might provide opportunities
to uncover this profoundly new realm of physics.

For what concerns the development
of a corresponding phenomenological program
the most promising opportunities for the
exploration of the possibility of genuinely quantum space-times
come from contexts in which there is no (classical) curvature.
If the interplay between general relativity and quantum mechanics
requires that at a fundamental level space-time is not classical,
then this should in particular be true for the space-times that
are perceived by our low-energy probes as flat and classical.
A space-time that appears to be Minkowski when probed 
by low-energy probes, would be perceived 
as ``quasi-Minkowski"~\cite{bignapap,gacgrf98}
(some appropriate ``quantum deformation" of Minkowski space-time)
by probes of higher energies.
[An example of such space-times
has been discussed in Ref.~\cite{gacmaj} and references therein.]

One of the effects that could characterize a quasi-Minkowski
space-time are quantum fluctuations of distances.
These are the focus of the present Article and will be discussed 
in detail in the following Sections.
Modern interferometers will be shown to provide exciting
opportunities to search for these distance fluctuations,
extending the preliminary indications of previous works on this
subject~\cite{gacgwi,polonpap,bignapap,nggwi,emngwi,fordgwi}.

Another important property of quasi-Minkowski space-times
is that the symmetries that characterize
a deformation of Minkowski space-time are of course
a deformed version (deformed algebras)
of the symmetries of Minkowski space-time.
This is extremely clear in the case of certain noncommutative
space-times~\cite{kpoin,gacmaj} whose symmetries are
properly described by Hopf algebras ({\it e.g.} 
the $\kappa$-Poincar\'{e} Hopf algebra, which reproduces the
ordinary Poincar\'{e} algebra only in the low-energy limit).
The realization that quantum gravity might lead
to deformed symmetries has led to renewed interest in certain
symmetry tests, as indirect tests of the short-distance structure
of space-time.

In particle physics the symmetries of Minkowski space-time
lead to the emergence of CPT symmetry.
CPT tests have been discussed in relation with
quantum gravity for more than 15 
years~\cite{ehns,hpcpt,elmn,kostpott,floreacpt,kaoka}.
The present upper limits on violations of CPT symmetry
(see, {\it e.g.}, Ref.~\cite{elmn})
have reached a level which is significant
for the study of the structure of space-time 
at Planckian distances. This is basically due to the fact
that limits on the neutral-kaon ``$\delta M_{K^0}/M_{K^0}$", 
one of the CPT-violation parameters
that can be introduced in the analysis of the neutral-kaon
system, have reached the level 
$\delta M_{K^0}/M_{K^0} \! < \! 10^{-19} \! \sim \! L_{p} M_{K^0}$.

There are two classes of sensitive tests of certain types
of deformations of Lorentz invariance that could 
be induced~\cite{grbgac,thooftlorentz,gampul} 
by non-trivial structure of space-time at Planckian distances.
One class of studies is based~\cite{grbgac,billetal} on the fact
that observations of the gamma rays we receive from distant
astrophysical sources allow to establish
that there is no anomalous effect
(within the achieved experimental accuracy)
leading to relative delays between
the times of arrival of simultaneously emitted photons.
The fact that these tests can be significant for quantum gravity
follows~\cite{polonpap} from the fact that the time-delay
sensitivity $\Delta T$ of the relevant experiments
is remarkably small as compared to the overall duration $T$ 
of the journey that the relevant particles make from their 
far away astrophysical emission point to the 
Earth: $\Delta T/T \sim 10^{-21} \sim L_p E$, where $E$ is
the energy of the particle.
A second class of tests of deformations of Lorentz invariance
is based~\cite{kifu,gactp,ita,aus}
on the fact that a deformation of Lorentz symmetries
would of course affect our estimates of the threshold energies
required for certain particle-production processes
(those thresholds are basically kinematical).
Recent experimental results 
concerning this second class of Lorentz-symmetry tests 
have led to some excitement.
In two energy regimes, photons around 10 TeV and cosmic rays
around $10^{20} eV$, certain puzzling data admit interpretation
as a manifestation of a departure from ordinary Lorentz
invariance~\cite{kifu,gactp,aus,colgla}.
As I shall emphasize again in Section 6,
the way in which these exciting results on ``threshold
anomalies"~\cite{gactp} have emerged
can provide encouragement for other quantum-gravity-phenomenology 
studies.

These four experimental programmes, the space-time-foam studies proposed in 
Ref.~\cite{gacgwi}, the CPT tests proposed in Ref.~\cite{grbgac}, the 
time-of-flight Lorentz-invariance tests proposed in Ref.~\cite{grbgac} and 
the threshold-energy Lorentz-invariance tests described in 
Ref.~\cite{gactp} (extending preliminary observations reported in 
Refs.~\cite{kifu,ita,aus}) are all we have at present as opportunities to 
explore experimentally effects associated with genuinely quantum 
space-times. The objective of quantum-gravity phenomenology, as defined 
earlier in this Article, is however even more general: one is hoping to 
gain insight on all aspects of the interplay between general relativity 
and quantum mechanics is of interest.
In this perspective there are at least two more classes of
observations which should be mentioned in this review Section.
The proposal put forward in Ref.~\cite{venegwi}
concerns possible tests (again using interferometers, but in
a way that is different from the one discussed in the present Article)
of the residual traces of some strong quantum-gravity effects
(specifically string-theory
effects~\cite{stringcosm})
which might have occurred in the early Universe.
The quantum-gravity effects considered in Ref.~\cite{venegwi}
are not of the type here defined as ``effects due to a genuinely
quantum space-time", since these effects
basically amount to the introduction of
new particles (the gravitons) in a classical
background space-time.

Some information on the behaviour of quantum mechanics
in presence of strong (but classical) gravitational
fields has been obtained in studies of
the quantum phases 
induced by large gravitational fields~\cite{cow,ahluexp}.
These experiments explore a very special regime
of quantum gravity, the one in which
the space-time aspects of the problem
can be analyzed within classical physics.
All aspects of space-time are treated classically
(one does not even introduce some new particles, {\it e.g.}
the gravitons, with space-time degrees of freedom)
but these experiments have provided~\cite{ahluexp} some insight 
on the role that the Equivalence Principle 
should have in quantum gravity.

\section{Planck-scale distance fluctuations could be
detected by modern interferometers}

Having briefly reviewed, in the preceding Section, the overall status
of quantum-gravity phenomenology, I now focus my attention
on the role that interferometers (and other types of detectors,
such as resonant bars) 
could have in the study of quantum-gravity-induced distance fluctuations.

A prediction of nearly
all approaches to the unification
of general relativity and quantum mechanics is that 
at very short distances the sharp
classical concept of space-time should give way 
to a somewhat ``fuzzy'' (or ``foamy'') 
picture (see, {\it e.g.}, Refs.~\cite{wheely,hawk,arsarea}),
but these new concepts are usually only discussed at a 
rather formal level.
If we are to test this prediction we must define space-time
fuzziness in physically meaningful (operative) terms.
Interferometers are the best tools for monitoring the distance
between test masses, and I propose as operative definition 
of the distance fluctuations that could be induced
by quantum gravity one which is expressed directly in terms
of strain noise in interferometers.\footnote{Since
modern interferometers were planned to look for classical
gravity waves (gravity waves are their sought ``signal"), 
it is reasonable to denominate as ``noise"
all test-mass-distance fluctuations 
that are not due to gravity waves.
I choose to adopt this terminology which reflects
the original objectives of modern interferometers,
even though this terminology is somewhat awkward
for the type of studies I am proposing 
in which interferometers would be used for searches
of quantum-gravity-induced distance fluctuations
(and therefore in these studies
quantum-gravity-induced distance fluctuations
would play the role of ``signal").}
In achieving their remarkable accuracy modern interferometers
must deal with several classical-physics strain noise 
sources ({\it e.g.}, thermal and seismic effects induce 
fluctuations in the relative positions of the test masses). 
Importantly, strain noise sources
associated with effects of ordinary quantum mechanics
are also significant for modern interferometers:
the combined minimization
of {\it photon shot noise} and {\it radiation pressure noise}
leads to a noise source which 
originates from ordinary quantum mechanics~\cite{saulson}.
The operative definition of fuzzy distance which I advocate
characterizes the corresponding quantum-gravity effects
as an additional source of strain noise.
A theory in which the concept of distance is 
fundamentally fuzzy in this 
operative sense would be such that
the read-out of an interferometer would still
be noisy (because of quantum-gravity effects)
even in the idealized limit in which all
classical-physics and ordinary-quantum-mechanics
noise sources are completely eliminated.
Just like the quantum properties of
the non-gravitational degrees of freedom of the apparatus
induce noise ({\it e.g.} the mentioned combination 
of {\it photon shot noise} and {\it radiation pressure noise})
it is of course plausible that noise be induced
by the quantum properties of
the gravitational degrees of freedom of the apparatus
({\it e.g.} the distances between the test masses).

Another simple way to discuss this operative definition
of distance fuzziness is the following. Let us assume that
we have established experimentally the exact dependence
on all relevant physical observables
of the total noise present in an interferometer.
The resulting strain noise spectrum will include terms
that are independent of both the Planck constant $\hbar$
and the gravitational constant $G$,
terms that depend either on $\hbar$ or on $G$,
and there could also be terms that depend on both $\hbar$ and $G$.
This last class of contributions to noise, 
depending on both $\hbar$ and $G$ (possibly on the particular 
combination of $\hbar$ and $G$ given by the Planck length $L_p$),
is here being defined as the quantum-gravity contribution
to noise.

This operative definition of quantum-gravity-induced
distance noise immediately confronts us with a potentially
serious challenge, which is the central challenge of
all quantum-gravity-phenomenology research lines:
if indeed quantum-gravity effects are proportional 
to (some power of) the Planck length $L_p$,
the smallness of $L_p$ will authomatically lead to
very small effects. However, 
modern interferometers
have a truly remarkable sensitivity to distance fluctuations
and it is actually not difficult to realize that this
sensitivity is potentially significant for the detection
of fluctuations occurring genuinely at the Planck scale.
In order to support this observation with a simple
intuitive argument let us consider the possibility that the 
distances $L$ betweeen the test masses of an interferometer be 
affected by Planck-length 
fluctuations of random-walk type
occurring at a rate of one per Planck time ($\sim 10^{-44} s$).
It is easy to show~\cite{gacgwi,polonpap,bignapap}
that such fluctuations would induce strain noise 
with power spectrum given by $L_p c L^{-2} f^{-2}$.
For $f \sim 100 Hz$ and $L\sim 1 Km$ 
(as for some modern interferometers) this corresponds 
to strain noise at the level $10^{-37} H\!z^{-1}$,
well within the reach of the sensitivity of
modern interferometers.

Fluctuations genuinely at the Planck scale
(the simple scheme I used to illustrate my point
involves Planck-length fluctuations occurring
at a rate of one per Planck time)
can lead to an effect that, while being very small in absolute
terms, is large enough for testing with modern interferometers.
This originates from the fact that random-walk fluctuations
do not fully average out. They have zero mean (in this sense
they do average out) but the associated standard deviation
grows with the time of observation (with the 
random-walk-characteristic $\sqrt{t}$ dependence which 
translates~\cite{polonpap,bignapap,rwold}
into the $f^{-2}$ dependence of the power spectrum).
A reasonable scale to characterize the time of observation
in interferometry is provided by $f^{-1}$ 
which, for $f \sim 100 H\!z$, is much larger than the Planck time.
[$(100 H\!z)^{-1}/10^{-44} s \sim 10^{40}$ and therefore
over a time of order $(100 H\!z)^{-1}$ the standard deviation
can become much greater than the Planck length.]

\section{Phenomenological description of space-time foam
and the sensitivity of planned experiments}

The example of random-walk fluctuations is quite interesting
since various quantum-gravity scenarios have 
random-walk elements~\cite{bignapap,fotinilee,ambjornrw}.
However, the random-walk case was here analyzed 
only as an example in which the classical space-time
picture breaks down on distance scales of order $L_p \sim 10^{-35}m$,
but the nature of this breaking is such that an interefometer
working at a few hundred $Hz$ is sensitive to
a collective effect of a very large number of minute 
fluctuations.\footnote{Here the analogy with the strategy adopted 
in proton-decay experiments is very direct.}
It may well be that the fluctuations induced by quantum gravity
are not of random-walk type, but it appears that interferometers
(and possibly resonant-bar detectors) should have significant
sensitivity to various scenarios in which the fluctuations 
average out only in the sense of the mean
and not in the sense of the standard deviation.

Having established that it is not preposterous to hope that
modern interferometers might have the capability to detect 
quantum-gravity-induced distance fluctuations,
it is now important for theorists to provide to the 
experimentalists  a detailed 
description of the fluctuations.
Unfortunately, the scarcity of experimental information on the
quantum-gravity realm has not yet allowed a 
proper ``selection process", so there are a large number of
quantum-gravity candidates. Moreover, even the two approaches whose
mathematical/logical consistency
has been already explored in some depth,
the one based on ``critical superstrings''~\cite{string1,string2}
and the one based on ``canonical/loop quantum
gravity''~\cite{cqgab,cqgcar,cqglee},
have not yet matured a satisfactory understanding
of their physical implications, 
such as the properties of space-time foam. 
In the few phenomenological programmes investigating
other quantum properties of 
space-time~\cite{ehns,hpcpt,elmn,kostpott,grbgac}
the difficulties deriving from the preliminary status 
of quantum-gravity theories have been 
circumvented by developing
direct phenomenological descriptions of the relevant phenomena.
I propose to apply the same strategy to the
description of the noise induced in interferometers
by quantum gravity.

As mentioned in Section~1,
my task is partly facilitated by the fact that
in order to guide interferometric studies
of foam it is only necessary to estimate
a relatively simple (single-variable) function:
the power spectrum $\rho_h(f)$ 
of the strain\footnote{Strain here has the standard
engineering definition $h \equiv \Delta L/L$ 
in terms of the displacement $\Delta L$ in a given
distance $L$.}
noise~\cite{saulson,rwold}.
In fact, the strain noise power spectrum,
through its dependence on the frequency $f$ at which 
observations are performed,
contains the most significant information
on the distance fluctuations, such as the mean square deviation
(which is given by the integral of the power spectrum over
the bandwidth of operation of the detector),
and is the quantity against which the observations are compared.

The quantum-gravity-induced strain noise should depend only
on the Planck length, 
the speed-of-light constant $c$ ($c \simeq 3 {\cdot} 10^8 m/s$),
and, perhaps, a length scale characterizing the properties
of the apparatus with respect to quantum gravity.
This renders dimensional-analysis arguments rather powerful:
if the analysis of a given quantum-gravity approach allowed
at least the identification of the qualitative 
nature of the distance fluctuations ({\it e.g.}, random walk)
then quite a lot of guidance could be provided to experimentalists
using simple dimensional arguments.
This is the central point being made in this Article and it
will be illustrated in a few examples relevant for quantum gravity,
also showing that in these examples the sensitivity of planned
interferometers can be significant.

The first qualitative picture that I want to analyze
dimensionally is one in which the foam-induced noise is white
(noise with constant, $f$-independent, power spectrum).
White noise is to be expected whenever~\cite{saulson,rwold}
the relevant stochastic phenomena
are such that there is no correlation between one
fluctuation and the next, an hypothesis which appears
to be rather plausible for the case of foam-induced
distance fluctuations.
The hypothesis that foam-induced noise be white is also
consistent with the intuition emerging from
analogies~\cite{garaythermal} between thermal environments
and the environment provided by foam as a (dynamical)
arena for physical processes.
According to these studies one can see foam-induced noise as
essentially analogous to thermal noise in various physical
contexts (such as electric circuits), 
which is indeed white whenever the bandwidth of interest is below 
some characteristic (resonant) frequency.
In the case of foam-induced noise the characteristic
frequency (which should be somewhere in the neighborhood of
the quantum-gravity frequency scale $c/L_p$)
would be much higher than the frequencies of operation
of our interferometers, and foam noise would be white
at those frequencies.

Within a white-noise model, by observing that the
strain noise power spectrum carries dimensions of $H\!z^{-1}$,
one is naturally led to the estimate 
\begin{equation}
 \rho_h(f) = {\rm constant} \sim  {L_p \over c} 
\sim 5 {\cdot} 10^{-44} H\!z^{-1} ~.
\label{white}
\end{equation}

I also observe that, since, as mentioned,
the frequencies we can access experimentally
are much smaller than $c/L_p$, 
white noise is actually the only admissable
structure for foam-induced strain noise within the hypothesis that
this noise be independent of the characteristics of the apparatus
which is used as a space-time probe.
In fact this hypothesis implies that $\rho_h$
can only depend on its argument $f$,
on the Planck length
and on the speed-of-light constant,
and therefore the most general low-frequency expansion
is of the type
\begin{equation}
 \rho_h(f) =
a_0 {L_p \over c} + a_1 \left( {L_p \over c} \right)^2 f
+ a_2 \left( {L_p \over c} \right)^3 f^2 + ...
\label{expans}
\end{equation}
where the $a_i$ are numerical coefficients
and all monomials of the type $f^{-|n|}$
were not included in the expansion because they would require
coefficients of the type $L_p^{-|n|+1}$ (which are inconsistent
with the fact that quantum-gravity effects must disappear in
the limit  $L_p \rightarrow 0$).
For $f \ll c/L_p$
the expansion (\ref{expans}) is well approximated by its first term, 
which corresponds to the dimensional estimate (\ref{white}).
From the point of view of experimental tests
it is also important to consider the value of the coefficient $a_0$,
{\it i.e.} to take into account the inherent uncertainty
associated with the dimensional estimate (\ref{white}).
In this type of studies
based on dimensional analysis,
the natural guess, which often turns out to be correct, 
is that coefficients such as $a_0$ are of order 1, 
but it is not uncommon to find
a disagreement between the dimensional estimate and the experimental
result of a few orders of magnitude.
In testing (\ref{white}) we shall therefore be looking for sensitivities
extending a few orders of magnitude below the $L_p/c$ level.

Since it does not involve any explicit dependence 
on the structure of the apparatus being
used to probe space-time, the estimate (\ref{white}) 
can be tested using any detector with sensitivity to distance strain,
such as interferometers and resonant-bar detectors.
Remarkably,
in spite of the smallness of the effects predicted,
these types of experiments are reaching such a high level
of sensitivity that (\ref{white}) is
going to be completely tested within a few years.

Denoting with $\rho_h^{TOT}$ the total
strain noise power spectrum observed by the experiments,
the present level of interferometric data
is best characterized
by the results obtained  
by the {\it 40-meter interferometer}~\cite{ligoprototype}
at Caltech and the {\it TAMA interferometer}~\cite{tama} at the
Mitaka campus of the Japanese National Astronomical Observatory,
both reaching $\rho_h^{TOT}$ of order $10^{-40} H\!z^{-1}$
(the lowest level has been achieved by {\it TAMA} around 
1$kH\!z$: $\rho_h^{TOT} \sim 3 {\cdot} 10^{-41} H\!z^{-1}$).
Even more remarkable is the present sensitivity
$\rho_h^{TOT} \simeq 5 {\cdot} 10^{-43}H\!z^{-1}$ 
of resonant-bar
detectors such as NAUTILUS~\cite{nautilus}
(which achieved it near $924 H\! z$).
This is already quite close to 
the estimate $L_p/c$ of (\ref{white}).
We are already probing a potentially interesting region and in
order to complete a satisfactory test of the estimate
(\ref{white}) we only need
to improve the sensitivity by a few orders of magnitude
(in order to exclude also the possibility that 
the coefficient $a_0$ be somewhat smaller than 1).

This will be accomplished in the near future.
Planned upgrades of the NAUTILUS resonant-bar detector
are expected~\cite{nautilus,micgwb}
to reach sensitivity
at the level $7 {\cdot} 10^{-45} H\!z^{-1}$.
The LIGO/VIRGO generation of 
interferometers~\cite{ligo,virgo} 
should achieve sensitivity of the 
order of $10^{-44} H\!z^{-1}$ within a year or two,
during its first phase of operation.
A few years later, with
the space interferometer LISA~\cite{lisa}
and especially with the ``advanced 
phase''~\cite{micgwb,ligo} of the LIGO/VIRGO interferometers,
another significant sensitivity improvement should be achieved:
according to recent estimates~\cite{ligo} it should be
possible to reach sensitivity levels in the neighborhood
of $10^{-48} H\!z^{-1}$, more than four orders of
magnitude below the $L_p/c$ estimate! 

This expected experimental progress is described in the figure
together with the $L_p/c$ white-noise level and the
analogous noise-level estimates that can be obtained
by assuming instead that the foam-induced noise be
of ``random-walk" type ({\it i.e.} with $f^{-2}$ frequency
dependence of the power spectrum~\cite{rwold}).
Through the example of random-walk noise the figure shows
that the sensitivity of modern interferometers is significant
also with respect to non-white models 
of foam-induced noise.
In the case in which the foam-induced distance fluctuations
are of random-walk type,
the corresponding strain noise
should necessarily depend on some
experiment-characteristic
length scale $\Lambda$
(differently from the case of $L_p$-linear white noise).
In fact, if random-walk noise depended only
on $f$, $c$, and $L_p$, the strain-noise 
spectrum would have to be
of the type $\rho_h \sim c L_p^{-1} f^{-2}$,
in contradiction with the fact that any Planck-scale effect
should vanish in the limit $L_p \rightarrow 0$.
A model with
random-walk strain noise 
linearly suppressed by the Planck
length would have to predict a power spectrum of the
form $\rho_h \sim c L_p f^{-2} \Lambda^{-2}$.
Our capability to test such a model
is to be described with the range of values of $\Lambda$
which we can exclude.
As shown in the figure,
for the $L_p$-linear random-walk-noise model
the excluded range of values of $\Lambda$ extends all the way
up to values of $\Lambda$ of the order of the length
of the arms of the interferometer.
In the random-walk
case we will soon even reach some sensitivity to
models with effects quadratically suppressed by the
Planck length; in fact,
as shown in the figure, the LISA
interferometer~\cite{lisa} will be able to test the possibility of
noise levels of the
type $\rho_h \sim c L_p^2 f^{-2} \Lambda^{-3}$,
at least in the case in which the scale $\Lambda$
is identified with the wavelength of the beam
used by LISA (which is, however, one of the smallest length
scale that characterize the experimental setup of LISA).
Since other quantum-gravity-motivated experimental programmes
can only achieve sensitivity to effects linear in the
Planck length~\cite{polonpap,ehns,hpcpt,elmn,kostpott,grbgac},
LISA's capability to reach some level 
of ``$L_p^2$ sensitivity"
will mark the beginning of another significant phase
in the search of quantum properties of space-time.

\begin{figure}[p] 
\begin{center} 
\epsfig{file=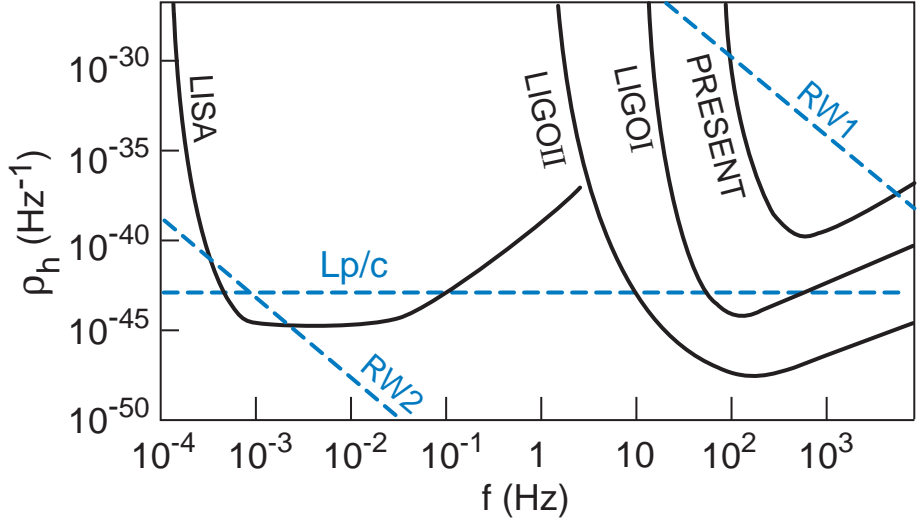,height=7truecm}\quad 
\end{center} 
\caption{\footnotesize A qualitative
(at best semi-quantitative) comparison
between the sensitivity of certain interferometers 
and the types of
strain noise power spectra ($\rho_h$) I am considering.
We expect significant progress from the level of sensitivity 
(``PRESENT'') of interferometers already in operation,
to the first phase of LIGO/VIRGO interferometers (``LIGOI''),
then to the second phase of LIGO/VIRGO (``LIGOII''),
and finally to LISA (``LISA'').
The horizontal line marks the noise level
corresponding to $L_p/c$.
The line ``RW1'' is representative of the random-walk scenario 
which is linearly suppressed by the Planck length and 
is proportional to the square of the inverse of the 
length of the arms of the interferometer.
In spite of the Planck-length suppression
the line RW1 is above (and therefore inconsistent with)
the noise levels achieved by ``PRESENT'' interferometers
(and by resonant-bar
detectors already in operation such as NAUTILUS~\cite{nautilus},
which achieved sensitivity $5 {\cdot} 10^{-43}H\!z^{-1}$,
near $924 H\! z$).
The figure also shows that with LISA 
we will achieve the capability to start the
exploration of some scenarios with quadratic
suppression by the Planck length:
the line ``RW2'' corresponds to random-walk noise
quadratically suppressed by the smallness
of the Planck length and proportional to the cubic power
of the inverse of the wavelength of the beam.
\label{fig:FIG1}}
\end{figure}

Having already discussed the possibility of white noise
and random-walk noise, let us consider just one more 
significant candidate: ``$f^{-1}$ noise".
It is in fact quite common in other noisy physical contexts
to find a noise contribution with $f^{-1}$ spectrum, 
and it appears reasonable at this early
stage of analysis to consider the possibility that also
quantum-gravity-induced strain noise might have this
characteristic behaviour.
Actually, some studies reported in Refs.~\cite{jare1,jare2}
appear to provide preliminary evidence of the possibility
that a minimum ingredient of quantum gravity is sufficient
for generating $f^{-1}$ noise.
In fact, I interpret the results obtained in Refs.~\cite{jare1,jare2}
as an indication that the minimum~\cite{thooftnew}
of distance fluctuations one would aspect in a quantum-gravity theory,
the fluctuations induced by the presence of gravitons,
is characterized by $f^{-1}$ noise.
Graviton effects would lead to $f^{-1}$ noise which is
quadratically suppressed by the Planck length 
($\rho_h \sim L_p^2 f^{-1} \Lambda^{-2}$),
and this is beyond the reach of forthcoming
experiments, unless the characteristic scale $\Lambda$
is to be identified with a rather short length
({\it e.g.}, the wavelength of the beam used in the interferometer).

Examples of $f^{-1}$ noise are not reported in the figure
(in order to maitain the number of lines in the figure
to a level that allows easy consultation);
however, the careful reader should realize that
the structure of $f^{-1}$ noise is somewhat intermediate
between the case of white ($f^{0}$) noise and
random-walk ($f^{-2}$) noise. It is therefore relatively
straightforward to deduce from the information provided in
the figure, and from the type of dimensional-analysis considerations
I reported above for white and random-walk noise,
that forthcoming experiments also have good sensitivity 
to some plausible candidates of foam-induced $f^{-1}$ noise.

\section{Some relevant issues for theoretical physics}

The main point of this Article, discussed in the preceding Section
(extending the analysis already reported in Ref.~\cite{gacgwi2}),
is that the phenomenology of quantum-gravity-induced distance
fluctuations can be quite effective starting from very
simple qualitative information on the nature of the fluctuations.
The two key ingredients are: (a) the general picture of the
fluctuation mechanism, and (b) the value of an experiment-characteristic
length scale $\Lambda$. Since at present even this qualitative 
information is not available in the various approaches to
quantum gravity, in the preceding Section I considered 3 plausible
scenarios for ingredient (a) (white noise, random-walk noise 
and $f^{-1}$ noise) and 2 plausible values of the
scale $\Lambda$ (the length of the arms of the interferometer
and the wavelength of the beam).
The number of scenarios to be considered will of course
be sharply reduced as soon as some clear theory indications
on the ingredients (a) and (b) are obtained.

While waiting for some quantum-gravity approaches
to reach this type of capability,
we can at least attempt to establish whether 
a given quantum-gravity approach would support some of the hypothesis
for the experiment-characteristic length scale $\Lambda$.
Two clear candidates are the length $L$ of the arms of the interferometer
and the wavelength $\lambda_0$ of the beam, but there are several
other length scales that characterize and interferometer,
and, even focusing only on $L$ and $\lambda_0$, it cannot be {\it a priori}
excluded that the value of $\Lambda$ (rather than being one of the simple
options $\Lambda \sim L$ or $\Lambda \sim \lambda_0$) be given by some 
combination of $L$ and $\lambda_0$, {\it e.g.} $\Lambda \sim L^2/ \lambda_0$.

It is also important to extend the analysis 
of the better-developed quantum-gravity approaches,
with the objective of establishing whether they
predict a genuinely quantum space-time, {\it i.e.}
establishing whether the emerging picture of space-time
requires something beyond the propagation of gravitational particles 
in an otherwise classical space-time.
The latest results on ``critical superstrings"
and ``loop quantum gravity" provide encouragement
for the idea of a genuinely quantum space-time.
These results suggest that at the quantum level
space-time has some elements of discreteness and/or
noncommutativity~\cite{spinnet,bilal,stringnonco}, and 
it appears unlikely that these 
features could be faithfully described by the propagation
of some new particles in an otherwise classical
space-time (continuous and commutative,
and such that there is no preferred frame for its description).

Another issue which could be discussed already at the
qualitative level concerns the key point
of the discussion in the preceding Section:~could
space-time fluctuations be of a type that averages out 
only in the sense of the mean
and not in the sense of the standard deviation?
The mentioned quantum-gravity scenarios with some 
random-walk elements~\cite{bignapap,fotinilee,ambjornrw}
provide encouragement for this possibility, but 
more work is needed, especially in order to develop
suitable effective theories.\footnote{In the formalisms
we ordinarily consider, such as the one of field theory,
effects on very short distances do not leave
any trace on larger scales. 
This is a key aspect of the renormalization procedure and 
of certain types of coarse graining.
For the fluctuations here considered this should correspond 
to the requirement that the mean vanishes and the standard 
deviation does not grow with time.}

One more point that deserves mention is the one concerning
whether or not there are some ``energy constraints''
that could be imposed on the structure of the quantum-gravity-induced
interferometric noise.
Energy considerations are clearly non-trivial in quantum gravity
(and general relativity) because one should not rely on
a given background space-time.
Effects describable in terms of particle propagation ({\it e.g.},
graviton propagation) in a given background space-time
can be easily analyzed from the point of view of energy considerations,
but, as mentioned, it appears likely~\cite{thooftnew} that quantum gravity
would predict also effects that cannot be described
in terms of particle propagation in a given background space-time.
For this second class of effects energy considerations are highly
nontrivial, for example it is easy to conceive~\cite{thooftnew}
space-time fluctuations that effectively ``carry" negative energy.
For pictures of space-time fluctuations that ``carry" positive energy
a strong constraint comes from the requirement not to overclose 
the Universe, a requirement which can be formalized through the
formula~\cite{micgwb}
\begin{equation}
{\rho_{gw} \over \rho_c} \simeq {10^{36} \over h_0^2}
\int_{f_{min}}^{f_{max}}
df \, \left( {f \over 1 {H\! z}}\right)^2 \rho_h(f)  ~,
\label{overclosing}
\end{equation}
where $\rho_{gw}$ is the energy density of
the stochastic background of gravity waves,
$\rho_c$ is the value of the critical energy
density for closing the Universe, and $h_0$
is a parameter which at present we can only constrain to 
the interval $0.50 < h_0 <0.85$ (reflecting the
experimental uncertainty in the Hubble constant).

Among the phenomenological models considered in the preceding Section,
this requirement can be applied
straightforwardly only to the one in which $f^{-1}$ noise originates
from the properties of gravitons (which indeed are effects associated
with the propation of new particles in a classical background
space-time). It is easy to see that the
fact that $f^{-1}$ noise tends quickly to $0$ at high frequencies
leads to authomatic compliance with the requirement.
For the other scenarios considered in the preceding Section,
which should not be associated with particle propagation
in a background space-time,
it is not clear whether one should be allowed to enforce
analogous energy requirements.
In addition to the mentioned fact that energy considerations
become highly nontrivial when there is no background space-time,
these requirements must also be treated prudently because
the proper way to understand quantum noise in interferometers
is that this noise is a property of the 
apparatus\footnote{Interferometric noise is always correctly
understood as a property of the interferometer with respect
to the relevant physical processes. This is true~\cite{saulson}
of the well-understood noise sources originating from
classical-physics phenomena ({\it e.g.}, thermal effects)
and from phenomena of ordinary (non-gravitational)
quantum mechanics (the mentioned combination
of photon shot noise and radiation pressure noise),
and it is here naturally assumed to be true of the
possible contribution to noise resulting from
quantum properties of space-time.}
(not a property of ``empty" space),
and therefore it is plausible that bounds based on energy 
considerations should involve some estimate of the energy 
that the fluctuations carry locally at the lab site 
(rather than considering some integration over the whole Universe).

The understanding of these delicate issues is however very important
for the development of the phenomenological approach here advocated.
It would not affect models with noise that decreases rapidly 
at high frequencies (just like in the case of  $f^{-1}$ noise,
also random-walk noise complies authomatically with requirements
of the type (\ref{overclosing})), but significant constraints
could emerge (if at all applicable) for models in which
there is no sharp decrease at high frequencies, such as
the white-noise model.

\section{Key challenges for the experimental programme}

As for the other research lines of quantum-gravity 
phenomenology,
these preliminary analyses of quantum-gravity-induced
distance fluctuations (and
the associated strain noise) should be interpreted as
an invitation to experimentalists to keep a vigilant
eye on possible anomalies that could be interpreted
as manifestions of the Planck-scale-structure of
space-time.
Quantum-gravity models are not yet ready for making definite
predictions, but one can examine some qualitative
aspects of a candidate new phenomenon and then make 
dimensional-analysis considerations in order to establish whether
relevant experiments have a chance to
uncover effects originating at Planckian distances.
As mentioned, at least in one case, the mentioned studies
of observed threshold anomalies, the fact that
experimentalists were alerted by this type of
considerations has led to exciting 
developments~\cite{kifu,gactp,ita,aus},
and it appears now reasonable to hope that
other quantum-gravity-phenomenology research lines 
might stumble upon analogous experimental results.

In the case of interferometers
(and resonant detectors) the task of keeping a vigilant
eye on possible quantum-gravity anomalies
is particularly challenging. In fact,
quantum gravity motivates the search for excess noise,
but excess noise of non-quantum-gravity origin is
very common in these experiments. Experimentalists make a
large effort of predicting all noise sources, but when the
machines finally are in operation it is quite natural
to find that one of the noise sources was underestimated.
Upon encountering excess noise the first natural guess is that
the excess be due to ordinary/conventional physics rather
than new physics.
In this respect one first point to be remarked is that
even just the determination of upper bounds on 
quantum-gravity-induced noise can be valuable at this
stage of development of quantum gravity. This research field 
has been for a long time a theoretical exploration of
a territory that was completely uncharted experimentally,
and it is therefore quite important that a few experiments are 
now providing at least some guidance in the form of upper limits 
naturally expressable in terms of the 
Planck length~\cite{ehns,grbgac,gacgwi}.
Conservative upper limits on quantum-gravity-induced 
excess noise can be set 
straigthforwardly~\cite{gacgwi,polonpap,bignapap}
by making conservative (lower bounds)
estimates of conventional
noise and observing that any quantum-gravity-induced 
noise could not exceed the difference between the 
noise observed and the corresponding conservative
estimate of conventional noise.

While upper limits are important, the possibility
of discovering quantum-gravity-induced 
noise would of course be more exciting.
As mentioned, the difficulties involved in
accurate predictions of noise levels of conventional
origin combine with the scarse information
on the structure to be expected for quantum-gravity 
noise in a way that renders such discoveries
very problematic; however, 
it is worth noticing that there are
certain characteristics of quantum-gravity-induced 
noise which could plausibly be used in order
to identify it.
For example, from the description given in the
preceding Sections it is clear that in an interferometer
quantum-gravity-induced noise might roughly look
like a stochastic background of gravity waves
with the important characteristic of the absence of
long-distance correlations.
Excess noise in the form of 
a stochastic background of gravity waves
is predicted also by other new-physics 
proposals\footnote{The early-Universe quantum-gravity
effects considered in Ref.~\cite{venegwi} would have 
eventually given rise to
a stochastic background of gravity waves, which
presently could play the role of noise in the observation of the 
gravity waves produced by other astrophysical phenomena.
The distance fluctuations induced by
the stochastic background of gravity waves
considered in Ref.~\cite{venegwi} would have
long distance correlations~\cite{micgwb}.}
but typically these other new-physics proposals
predict long-distance correlation~\cite{venegwi,micgwb}.
This important difference could be used to distinguish 
experimentally between noise induced by genuinely quantum
properties of space-time
and other excess-noise new-physics proposals.

\section{Closing remarks}

Encouraged by the exciting developments that have recently
emerged in another quantum-gravity-phenomenology research 
line~\cite{kifu,gactp,ita,aus}, I have here outlined 
a phenomenological approach to the description
of quantum-gravity-induced noise.
Based on the information contained in the figure
it appears that forthcoming experiments 
will start constraining quite severely model-building
for quantum-gravity-induced noise.

These opportunities are directly associated with the fact
that interferometers are preparing to reach sensitivity
at or below $10^{-44} H\!z^{-1}$ over a relatively wide
(combining LIGO/VIRGO and LISA)
range of frequencies.
It is quite amusing to notice that experimentalists
have been preparing for these sensitivity levels
in response to classical-physics thoretical
studies showing that the strain noise power spectrum should
be reduced at or below the level $10^{-44} H\!z^{-1}$
in certain frequency windows in order to allow 
the discovery of classical gravity waves.
It is a remarkable numerical accident that the result of these
classical-physics studies, involving several length scales
such as the distance between the Earth and potential sources
of gravity waves, has pointed us toward a sensitivity level
which I here observed to be also naturally described
in terms of the intrinsically quantum scale $L_p/c$. 


\newpage
\baselineskip 12pt plus .5pt minus .5pt


\begin{thebibliography}{99}

\bibitem{cow} R.~Colella, A.W.~Overhauser and S.A.~Werner,
Phys.~Rev.~Lett.~34 (1975) 1472.

\bibitem{ehns} J.~Ellis, J.S.~Hagelin, D.V.~Nanopoulos
and M.~Srednicki,
Nucl.~Phys.~B241 (1984) 381.

\bibitem{venegwi} R. Brustein, M. Gasperini, M. Giovannini
and G. Veneziano, Phys.~Lett.~B361 (1995) 45.

\bibitem{ahluexp}
D.V.~Ahluwalia, 
Mod.~Phys.~Lett.~A13 (1998) 1393.
 
\bibitem{grbgac} G. Amelino-Camelia, J. Ellis, N.E. Mavromatos, 
D.V. Nanopoulos and S. Sarkar, 
astro-ph/9712103,
Nature 393 (1998) 763.

\bibitem{gacgwi} G. Amelino-Camelia, 
gr-qc/9808029,
Nature 398 (1999) 216.

\bibitem{kifu} T.~Kifune,
Astrophys.~J.~Lett.~518 (1999) L21.

\bibitem{gactp} G.~Amelino-Camelia and T.~Piran, astro-ph/0008107.

\bibitem{polonpap} G.~Amelino-Camelia,
{\it Are we at the dawn of quantum-gravity phenomenology?},
Lect.~Notes~Phys.~541 (2000) 1-49.
      
\bibitem{grf00} G.~Amelino-Camelia, ``Planck-length phenomenology'',
gr-qc/0008010.

\bibitem{bignapap} G. Amelino-Camelia, 
Phys.~Rev.~D62 (2000) 024015.

\bibitem{nggwi} Y.J.~Ng and H.~van Dam,
Found.~Phys.~30 (2000) 795.
 
\bibitem{emngwi}
A.~Campbell-Smith, J.~Ellis, N.E.~Mavromatos 
and D.V.~Nanopoulos,
Phys.~Lett.~B466 (1999) 11. 

\bibitem{fordgwi} Hong-wei Yu and L.H.~Ford,
gr-qc/9907037.

\bibitem{gacgwi2} G.~Amelino-Camelia,
{\it A phenomenological description of space-time noise in 
quantum gravity}, Nature (2001) in press.

\bibitem{saulson} Saulson P.R., {\it Fundamentals of interferometric 
gravitational wave detectors} (World Scientific, Singapore, 1994).

\bibitem{rwold} Radeka V.,
{\it Low-Noise Techniques in Detectors},
Ann.~Rev.~Nucl.~Part.~Sci.~38 (1988) 217.

\bibitem{thooftnew} G.~'t Hooft,
Class.~Quantum Grav.~16 (1999) 3263. 

\bibitem{ahlu94grf} D.V.~Ahluwalia, 
Phys.~Lett.~B339 (1994) 301.

\bibitem{gacgrf98} G. Amelino-Camelia, 
Mod.~Phys.~Lett.~A13 (1998) 1319.

\bibitem{spinnet} C.~Rovelli and L.~Smolin,
Phys.~Rev.~D52 (1995) 5743.

\bibitem{kpoin} 
J.~Lukierski, A.~Nowicki and H.~Ruegg, 
Ann.~Phys.~243 (1995) 90.

\bibitem{gacmaj} 
G.~Amelino-Camelia, J.~Lukierski and A.~Nowicki, 
Int.~J.~Mod. Phys.~A14 (1999) 4575;
G.~Amelino-Camelia and S.~Majid,
Int.~J.~Mod.~Phys.~A15 (2000) 4301.

\bibitem{bilal} A.~Bilal,
Fortsch.~Phys.~47 (1999) 5.

\bibitem{stringnonco} N.~Seiberg and E.~Witten,
hep-th/9908142, JHEP 9909 (1999) 032.

\bibitem{hpcpt} P.~Huet and M.E.~Peskin,
Nucl.~Phys.~B434 (1995) 3 .

\bibitem{elmn} J. Ellis, J. Lopez, 
N. Mavromatos, D. Nanopoulos and CPLEAR Collaboration,
Phys.~Lett.~B364 (1995) 239 .

\bibitem{kostpott} V.A.~Kostelecky and R.~Potting,
Phys.~Lett.~B381 (1996) 89.

\bibitem{floreacpt} F.~Benatti and R.~Floreanini, 
Nucl.~Phys.~B {\bf 488}, 335 (1997) .

\bibitem{kaoka} G.~Amelino-Camelia and F.~Buccella,
Mod.~Phys.~Lett.~A15 (2000) 2119.

\bibitem{thooftlorentz}
G.~`t Hooft,
Class.~Quant.~Grav.~13 (1996) 1023.

\bibitem{gampul} R.~Gambini and J.~Pullin,
Phys.~Rev.~D59 (1999) 124021.

\bibitem{billetal} B.E.~Schaefer,
Phys.~Rev~Lett.~82 (1999) 4964;
S.D. Biller {\it et al},
Phys.~Rev.~Lett.~83 (1999) 2108. 

\bibitem{ita} 
R.~Aloisio, P.~Blasi, P.L.~Ghia and A.F.~Grillo,
Phys.~Rev.~D62 (2000) 053010.

\bibitem{aus} R.J.~Protheroe and H.~Meyer,
Phys.~Lett.~B493 (2000) 1.

\bibitem{colgla} S.~Coleman, S.L.~Glashow,
Phys.~Rev.~D59 (1999) 116008.

\bibitem{stringcosm} G.~Veneziano, Phys.~Lett.~B265 (1991) 287;
M.~Gasperini and G.~Veneziano, Astropart.~Phys.~1 (1993) 317.

\bibitem{wheely} J.A.~Wheeler, in {\it Relativity, groups and topology},
(eds.~B.S.~De Witt \& C.M.~De Witt) 
(Gordon and Breach, New York, 1963).

\bibitem{hawk} S.W.~Hawking, {\it Spacetime foam},
Nuc.~Phys.~B144 (1978) 349.

\bibitem{arsarea} A.~Ashtekar, C.~Rovelli and L.~Smolin,
Phys.~Rev.~Lett.~69 (1992) 237.

\bibitem{fotinilee} F.~Markopoulou and L.~Smolin,
Phys.~Rev.~D58 (1998) 084033.

\bibitem{ambjornrw} R.~Loll, J.~Ambjorn and K.N.~Anagnostopoulos,
Nucl.~Phys.~Proc.~Suppl.~88 (2000) 241.

\bibitem{string1} M.B.~Green,
J.H.~Schwarz, \& E.~Witten, {\it Superstring theory}
(Cambridge Univ.~Press, Cambridge, 1987).

\bibitem{string2} J.~Polchinski, {\it String theory}
(Cambridge Univ.~Press, Cambridge, 1998).

\bibitem{cqgab} A.~Ashtekar, 
{\it Quantum mechanics of geometry}, gr-qc/9901023.

\bibitem{cqgcar} M.~Gaul and C.~Rovelli,
{\it Loop Quantum Gravity and the Meaning of Diffeomorphism Invariance},
Lect.~Notes~Phys.~541 (2000) 277-324.

\bibitem{cqglee} L.~Smolin,
{\it The new universe around the next corner},
Physics World 12 (1999) 79-84.

\bibitem{garaythermal} L.J.~Garay, {\it Space-time foam 
as a quantum thermal bath},
Phys.~Rev.~Lett.~80 (1998) 2508.

\bibitem{ligoprototype} A.~Abramovici {\it et al}, 
Phys.~Lett.~A218 (1996) 157.

\bibitem{tama} Updated information
on the progress of observations performed by the
TAMA interferometer can be found
at the WWW site http://tamago.mtk.nao.ac.jp/.

\bibitem{nautilus} Astone P.~{\it et al},
Phys.~Lett.~B385 (1996) 421.

\bibitem{micgwb} M.~Maggiore,
Physics Reports 331 (2000) 283.

\bibitem{ligo} A.~Abramovici {\it et al}, 
{\it LIGO: The Laser Interferometer Gravitational-Wave
Observatory},
Science 256 (1992) 325-333.
(Updated information
on expected sensitivity of an advanced phase
of the LIGO interferometer can be found
at WWW site http://www.ligo.caltech.edu/~ligo2/.)
 
\bibitem{virgo} B.~Caron {\it et al},
{\it The Virgo interferometer},
Class.~Quantum Grav.~14 (1997) 1461-1469.
(Details on the sensitivity objectives of VIRGO can be
found at the WWW site http://www.virgo.infn.it/.)

\bibitem{lisa} K.~Danzmann,
{\it LISA: Laser interferometer space antenna for 
gravitational wave measurements},
Class.~Quantum Grav.~13 (1996) A247.

\bibitem{jare1} M.-T.~Jaekel and S.~Reynaud,
Europhys.~Lett.~{\bf 13} (1990) 301.

\bibitem{jare2} M.-T.~Jaekel and S.~Reynaud,
Phys.~Lett.~{\bf B185} (1994) 143.

\end{thebibliography}
\end{document}